\documentclass{article}
\usepackage{spconf,amsmath,graphicx,hyperref}
\usepackage[numbers]{natbib}
\usepackage{booktabs,tabularx,array}
\newcolumntype{Y}{>{\raggedright\arraybackslash}X} 

\title{Semantic Visually-Guided Acoustic Highlighting with Large Vision-Language Models}
\name{Junhua Huang, Chao Huang, Chenliang Xu}
\address{University of Rochester}
\begin{document}
%
\maketitle
\begin{abstract}
Balancing dialogue, music, and sound effects with accompanying video is crucial for immersive storytelling, yet current audio mixing workflows remain largely manual and labor-intensive. While recent advancements have pioneered the visually-guided acoustic highlighting task, implicitly re-balancing audio sources with multi-modal guidance, what visual aspects serve as effective conditioning remains largely unknown. 
We address this gap with a systematic study of whether \emph{deep video understanding} improves audio remixing. Using textual descriptions as proxy for visual analysis, we ask large vision-language models (LVLMs) to extract six aspect types, including object/character appearance, emotion, camera focus, tone, scene background, and inferred sound-related cues. In extensive experiments, \emph{camera focus}, \emph{tone}and \emph{scene background} consistently yield the highest gains in perceptual mix quality over state-of-the-art baselines. Our findings (i) identify which visual–semantic cues most strongly support coherent, visually aligned remixing, and (ii) chart a practical path toward automating cinema-grade sound design via lightweight, LVLM-derived guidance.
\end{abstract}
%

\section{Introduction}
\label{sec:intro}

Film, television, and games rely on a tightly choreographed duet of vision and sound, yet much of today’s stem mixing remains manual: trade guidance estimates \emph{1 engineer-hour per finished minute} for dialogue-centric content, with complex projects reaching \emph{5–10 hours per finished minute}~\cite{iZotope2018Workflows}. Late picture-lock changes trigger additional passes, and mismatches between visual focus and loudness undermine immersion and accessibility. An automated system that rebalances stems \emph{in line with the video} could reduce costs and broaden access to high-quality mixes.

Audio-only pipelines (separate, then heuristically remix) ignore the image and can misalign with visual intent. Visually-Guided Acoustic Highlighting (VisAH)\cite{huang2025visah} re-framed remixing as visually guided audio-to-audio translation and showed that \emph{text} descriptions of the video can rival learned video features for guidance. However, VisAH operated at the shot/segment level and did not disentangle \emph{which} visual semantics most strongly predict desirable mix adjustments; the relation between specific visual elements and mix composition remains under-explored.

To address this, we look back to established practice and audiovisual literature: mixers routinely prioritize on-screen sources and preserve dialogue intelligibility via ducking and related controls~\cite{iZotope2018Workflows}. Technical studies further show that visual evidence identifies sound sources and supports source-selective level control~\cite{huang2025visah}. These observations suggest that concrete cues, such as camera focus, scene/background context, object/character identity, affect, and visible cause–of-sound should be especially predictive for remixing.

Modern LVLMs can surface such cues in text form without heavyweight video encoders: they localize and ground entities~\cite{Peng2023Kosmos2}, align with human gaze/attention~\cite{Yan2024VoilaA}, and reason about hands and object interactions~\cite{Bao2024HandsOnVLM}. Motivated by this, we study \emph{LVLM-conditioned remixing}: a frozen LVLM produces concise textual descriptors as proxies for frame-level semantics, and we test targeted subsets as conditioning signals.  We find that \textit{camera focus} and \textit{scene}  and \textit{Sound Sources} are the most influential components, and that with simple prompt refinements, using LVLM-extracted cues yields state-of-the-art gains over audio-only and prior multi-modal baselines.

\begin{figure*}[t]
  \centering
  \includegraphics[width=.8\textwidth]{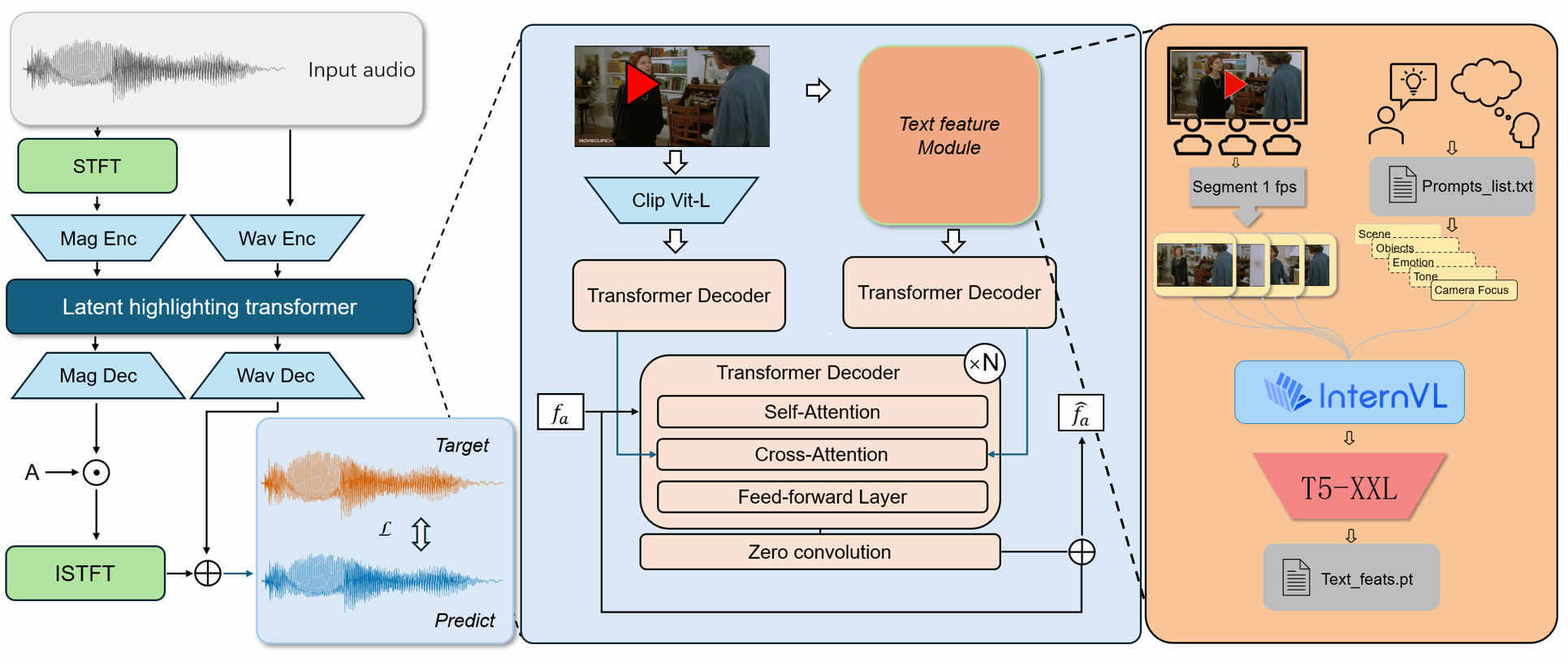} 
  \caption{Overview identical to VisAH~\cite{huang2025visah} except the \emph{text feature module} (orange) feeding the context encoder.}
  \label{fig:pipeline}
  \vspace{-1.0ex} 
\end{figure*}

\section{Related Work}
\label{sec:related}

\noindent\textbf{From separation to remixing.}
Early ``auto-mix'' pipelines first estimate stems via audio-only source separation and then apply heuristic gain curves; widely used open-source systems include Open-Unmix~\cite{Stoter2019OpenUnmix} and Demucs~\cite{Defossez2019Demucs}. To exploit video information, audio--visual correspondence methods align visible events with sound.
Closest to our setting, \emph{Learning to Highlight Audio by Watching Movies} (VisAH) formalizes visually guided acoustic highlighting as \emph{implicit remixing} rather than explicit separation: it unifies source separation and gain re-balancing by predicting a single remix mask, and trains an end-to-end audio-to-audio framework on the synthetic \textsc{MuddyMix} corpus, outperforming audio-only baselines. We likewise follow this end-to-end \emph{remixing} formulation and ask which visual/text cues most effectively steer time-varying gain control.

\noindent\textbf{LVLMs as fine-grained conditioning signals.}
Recent large vision--language models (LVLMs) provide grounded, frame-level semantics aligned with editorial intent (e.g., viewer gaze/attention~\cite{Yan2024VoilaA} and text--region grounding~\cite{Peng2023Kosmos2}). Building on these advances, we adopt textual proxies from a frozen LVLM (keyframes $\rightarrow$ captions $\rightarrow$ text embeddings) to condition a lightweight gain controller, and we systematically ablate six semantic aspects: appearance, emotion, tone, camera focus, scene background, and inferred sound-related cues, to identify which cues most improve remixing quality.

\section{Approach}
\subsection{Task Overview}
\label{sec:task}

We study \emph{visually guided acoustic highlighting (VGAH)}: given a short video clip $V$ and its rough or poorly balanced composite mix $A$, produce a rebalanced soundtrack $\hat{A}$ that preserves content (dialogue, music, Foley) while adjusting relative loudness to follow on-screen narrative salience. Prior audio-only pipelines and VisAH variants conditioned on undifferentiated video features at the shot/segment level, leaving underexplored \emph{which specific visual semantics} most strongly predict desirable mix adjustments.
To address this, we propose SemMix, studying what concrete visual signals (\underline{Sem}antics) trigger better re-balancing effects (\underline{Mix}ing). 

\subsection{Pipeline Overview}
Our architecture follows VisAH~\cite{huang2025visah}: a dual-branch audio backbone (time and frequency), a context encoder, a latent highlighting transformer, and a mask-based decoder with iSTFT.
Our focus here is the prompt pathway that forms the textual context (Fig.~\ref{fig:pipeline}).

\subsection{SemMix: What Visual Semantics Matter?}
We ground our design in strands of previous work. From film sound practice, classic sources emphasize that effective remixing hinges on a small set of perceivable cues: emotional performance on screen, concrete sound-producing objects, spatial/temporal setting, overall visual mood, visible sound sources and what the camera emphasizes—because these cues drive dialogue presence, Foley selection, ambience, and editorial salience \cite{Chion1994AudioVision,Holman2010SoundFilmTV}. From the LVLM literature, studies show that models hallucinate when instructions are ambiguous or invite off-screen inference; prompting them to remain within visible evidence and to abstain when uncertain mitigates such errors \cite{rohrbach-etal-2018-object,Huh2024VizWizLF}. Synthesizing these observations, we developed six topics that capture the key decision levers of film sound: diegetic anchors that guide mixing, scene / time / tone cues that shape ambience, and focus that aligns audiovisual attention while avoiding backstory or off-screen inference \cite{Chion1994AudioVision,ThomDesigningForSound}.

\begin{itemize}
\vspace{-0.1in}
  \item \textbf{Emotion (Actors)}: dominant on-screen emotions; note a secondary cue only if clearly visible.
  \vspace{-0.1in}
  \item \textbf{Objects (Salient)}: brief inventory of eye-catching, sound-relevant \emph{visible} items (nouns).
  \vspace{-0.1in}
  \item \textbf{Scene (Setting \& Time)}: where/when plus lighting and crowding inferred from what is on screen.
  \vspace{-0.1in}
  \item \textbf{Tone (Color \& Mood)}: palette and lighting style that shape ambience/bed choice.
  \vspace{-0.1in}
  \item \textbf{Sound Sources (Visible)}: on-screen elements likely to emit sound (diegetic anchors).
  \vspace{-0.1in}
  \item \textbf{Camera Focus (Salience)}: intended subject of attention and cues that visibly signal that focus.
\end{itemize}

\begin{table*}[!ht]
  \caption{Main comparison: The best results are in \textbf{bold}, the second best are \underline{underlined}. Metrics are $\times 100$. Percentages in parentheses indicate improvement vs \textit{Poorly Mixed Input} (lower is better).}
  \label{tab:main}
  \centering
  \footnotesize
  \setlength{\tabcolsep}{6pt}
  \renewcommand{\arraystretch}{1.12}
  \begin{tabularx}{0.9\textwidth}{Y c c c c c}
    \toprule
    Method & MAG $\downarrow$ & ENV $\downarrow$ & KLD $\downarrow$ & $\Delta$IB $\downarrow$ & W\mbox{-}dis $\downarrow$ \\
    \midrule
    \textit{Poorly Mixed Input} & 22.69 & 6.30 & 20.61 & 1.52 & 1.94 \\
    DnRv3~\cite{watcharasupat2024dnrv3}+CDX~\cite{Uhlich2024CDX}
      & 26.32~($-16\%$) & 7.62~($-21\%$) & 15.87~($+23\%$) & $1.78$~($-17\%$) & 2.84~($-46\%$) \\
    Learn2Remix~\cite{yang2022learn2remix}
      & 19.07~($+16\%$) & 4.16~($+34\%$) & 61.76~($-200\%$) & 8.27~($-444\%$) & \underline{1.20~($+38\%$)} \\
    LCE\textendash SepReformer~\cite{jiang2024lce}
      & 17.18~($+24\%$) & 4.28~($+32\%$) & 30.99~($-50\%$) & 1.88~($-24\%$) & 1.28~($+34\%$) \\
    VisAH~\cite{huang2025visah}
      & \underline{10.08~($+56\%$)} & \underline{3.43~($+46\%$)} & \underline{11.01~($+47\%$)} & \textbf{0.80~($+47\%$)} & \textbf{0.79~($+59\%$)} \\
    \textbf{SemMix-Camera Focus}
      & \textbf{9.99~($+56\%$)} & \textbf{3.41~($+46\%$)} & \textbf{10.95~($+47\%$)} & \underline{0.87~($+43\%$)} & \textbf{0.79~($+59\%$)} \\
    \bottomrule
  \end{tabularx}
\end{table*}

\vspace{-0.05in}
\subsection{Prompt Versions.}
We distill two complementary families whose inspirations differ. \emph{Focused} prompts add explicit constraints and, where helpful, brief illustrative patterns to keep the model strictly grounded in what is visible and to permit abstention when evidence is insufficient \cite{Huh2024VizWizLF}. \emph{Minimal} prompts keep instructions as short as possible to test whether concise guidance suffices and to reduce prompt overhead; this parsimony is motivated by findings that longer contexts dilute attention to salient details and can degrade faithfulness \cite{liu-etal-2024-lost}. Thus, families target different risks—underspecification versus overspecification—and are not overlapping in intent.

To make the contrast concrete with two examples: for \emph{Sound Sources}, in a quiet office shot where only a ceiling fan is plainly shown, the Focused version requires staying on-screen and returning \emph{none} when no source is evident, yielding a single, grounded cue (e.g., fan hum); the Minimal version simply lists likely on-screen sources, which keeps output brief but omits the explicit abstention rule. For \emph{Camera Focus}, in a tight close-up of a hand on a doorknob with the hallway blurred, the Focused version names the intended subject \emph{and} cites the visible technique enforcing that focus (e.g., shallow depth of field), whereas the Minimal version reports only the subject. Comparisons between the two versions are deferred to Sec.~5.2 under the same evaluation metrics as our prior VisAH setup.

\section{Experiment}
\subsection{Experimental Setting}
We follow the VisAH setup~\cite{huang2025visah} and modify only the text pathway. All data splits, audio preprocessing (44.1\,kHz stereo to mono), STFT parameters, backbones, and default schedules are identical to VisAH unless stated otherwise. Text embeddings use a channel dimension of $C_{\text{text}}=4096$ and enter through the same conditioning interface. Models are trained with Adam (learning rate $10^{-4}$), batch size $12$ per GPU, for $200$ epochs. 

Training is performed on $2\times$ RTX\,4090 GPUs; a full run requires approximately $21$ hours.

\vspace{0.5ex}
\subsection{Evaluation Metrics}
To attribute improvements solely to prompting, we adopt VisAH’s metric suite without modification. \emph{MAG}$\downarrow$ quantifies spectral accuracy via the L1 distance between the magnitude STFTs, while \emph{ENV}$\downarrow$ evaluates the dynamics in the time domain using envelopes derived from the analytical signal. For semantics, we compute \emph{KLD}$\downarrow$ between PaSST-based event distributions and report the ImageBind-based gap $\Delta\mathrm{IB}=\mathrm{IB}(V,A)-\mathrm{IB}(V,\hat{A})$, where smaller values indicate closer audio-video correspondence. Finally, \emph{W-dis}$\downarrow$ measures the cost to align predicted versus reference loudness trajectories (speech/music/effects), capturing timing and balance over time.

\subsection{Baselines and Prior Work}
We adopt the exact baseline suite used by \textbf{VisAH} and follow its evaluation split; we also rerun the released configurations to verify that our reproduced numbers match within typical tolerance \cite{huang2025visah}. The baselines are the following.

\begin{itemize}
    \vspace{-0.05in}
\item \textbf{Poorly-Mixed Input.} The manually constructed poorly mixed input from the dataset creation process was used as a reference point.
\vspace{-0.05in}
\item \textbf{DnRv3~\cite{watcharasupat2024dnrv3}\,+\,CDX~\cite{Uhlich2024CDX}.} Apply the DnR\,v3 separator to obtain three stems (speech, music, SFX), then sample loudness per stem from the CDX distributions and remix to produce the output.
\vspace{-0.05in}
\item \textbf{Learn2Remix~\cite{yang2022learn2remix}} An end-to-end neural remixer; following VisAH, we replace the original Conv-TasNet backbone with \emph{SepReformer} for a stronger separator and train on our data.
\vspace{-0.05in}
\item \textbf{Listen, Chat, and Edit (LCE)~\cite{jiang2024lce}} A text-guided mixture editor; as in VisAH, explicit instructions are unavailable at test time, so we provide text captions as guidance. For parity, we substitute the SoundEditor backbone with \emph{SepReformer}.
\end{itemize}

We treat VisAH itself as the prior SOTA that fuses raw vision and text; our contribution isolates the benefit of richer, prompt-engineered captions (with InternVL-family captioners \cite{Chen2024InternVL}) while leaving the audio stack unchanged. Table~\ref{tab:main} summarizes the main comparison in \textsc{MuddyMix};\textbf{ Our SemMix method uses 18.94M fewer parameters that reach the previous SOTA level performance with a reduction of \emph{9.0\% } in the size of the model.}

\section{Analysis}

\begin{figure}[!t]
  \centering
  \includegraphics[width=\columnwidth]{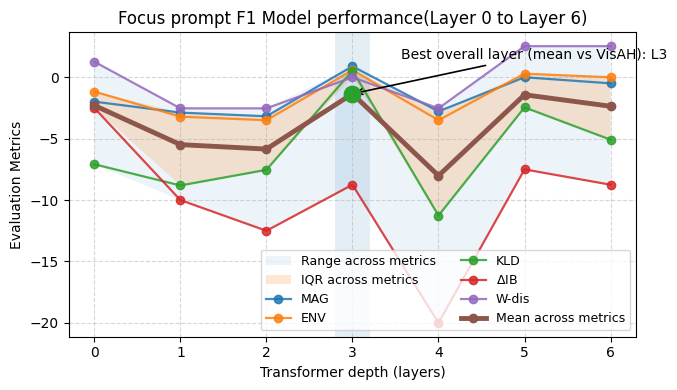}
  \caption{Model performance using focused prompt \textbf{Camera Focus} from layer 0 to layer 6. shaded bands show cross-metric spread (min--max and IQR). The mean curve peaks at \textbf{$L{=}3$}; $L{=}5/6$ offer mild W-dis polish.}
  \label{fig:depth_trend}
\end{figure}

\subsection{Prompt Comparison.}
We benchmark six prompt designs under an identical setup (Table~\ref{tab:prompt_comparison}). Performance varies significantly between prompts. \textbf{Objects} degrades all metrics, suggesting that listing salient items biases the model toward visually striking but acoustically irrelevant details. \textbf{Scene} ,\textbf{Sound Source} and \textbf{Tone} yield small but consistent magnitude (MAG) reductions, while \textbf{Scene } also improves KLD, indicating that the coarse spatio-temporal context helps align the separation targets. The strongest gains come from \textbf{Camera Focus}, which reduces MAG by 3.2\%, ENV by 3.4\%, and KLD by 6.0\% relative to the baseline; concentrating the description on what the camera makes primary appears to filter out distractors and emphasize sources most likely to dominate the mix. \textbf{Emotion} offers marginal MAG improvement but harms ENV/KLD, implying affect cues are weakly coupled to source structure. Overall, \textbf{Camera Focus} is the leading prompt, followed by \textbf{Sound Source} and \textbf{Scene}.

\begin{table}[!t]
\caption{Overall prompt comparison. The best result per column is in \textbf{bold}; values better than VisAH are \underline{underlined}. Improvements are relative to VisAH (baseline). Lower is better.}
\label{tab:prompt_comparison}
\centering
\scriptsize
\setlength{\tabcolsep}{3pt}
\renewcommand{\arraystretch}{1.05}
\begin{tabular}{l c c c c}
  \toprule
  Prompt & \#Layers & MAG $\downarrow$ & ENV $\downarrow$ & KLD $\downarrow$ \\
  \midrule
  VisAH (baseline)         & 3 & 10.34 (0\%)           & 3.51 (0\%)           & 11.75 (0\%) \\
  Emotion (Actors)         & 3 & 10.32 (+0.2\%)        & 3.54 ($-$0.9\%)      & 11.97 ($-$1.9\%) \\
  Objects (Salient)        & 3 & 10.73 ($-$3.8\%)      & 3.68 ($-$4.8\%)      & 12.07 ($-$2.7\%) \\
  Scene (Setting \& Time)  & 3 & \underline{10.28 (+0.6\%)} & \underline{3.50 (+0.3\%)} & \underline{11.53 (+1.9\%)} \\
  Tone (Color \& Mood)     & 3 & \underline{10.24 (+1.0\%)} & 3.51 (0.0\%)        & 12.10 ($-$3.0\%) \\
  Sound Sources (Visible)  & 3 & \underline{10.29 (+0.5\%)} & \underline{3.49 (+0.6\%)} & \underline{11.51 (+2.0\%)} \\
  Camera Focus (Salience)  & 3 & \textbf{\underline{10.01 (+3.2\%)}} & \textbf{\underline{3.39 (+3.4\%)}} & \textbf{\underline{11.05 (+6.0\%)}} \\
  \bottomrule
\end{tabular}
\end{table}

\subsection{Focused vs.\ Minimal Prompt Sets.}
Given the page limit, we provide a compact comparison between the two families. Under the same evaluation setup (Table~\ref{tab:prompt_sets_compact}), the \textbf{Focused Prompt Set } generally matches or slightly exceeds the \textbf{Minimal Prompt Set } on our metrics, suggesting that brief but explicit cues (e.g., camera focus and scene context) help align visual evidence with the remixing objective. This observation is consistent with previous findings that model outputs are highly sensitive to prompt formulation and calibration \cite{zhao2021calibrate}, that light structure / guidance improves fidelity in tasks where the target is not directly observable \cite{wang2023selfconsistency}.

\begin{table}[t!]
  \caption{Focused vs.\ Minimal prompt sets (lower is better). Replace dashes with your measured values or relative deltas.}
  \label{tab:prompt_sets_compact}
  \centering
  \footnotesize
  \setlength{\tabcolsep}{6pt}
  \renewcommand{\arraystretch}{1.1}
  \begin{tabular}{lccc}
    \toprule
    Prompt Set & MAG $\downarrow$ & KLD $\downarrow$ & W-dis $\downarrow$ \\
    \midrule
    Focused  & 9.99 & 10.95 & 0.79 \\
    Minimal  & 10.29 & 12.36 & 0.78 \\
    \bottomrule
  \end{tabular}
\end{table}

\subsection{Transformer Depth.}
Following the depth settings explored in \cite{huang2025visah} (0, 3, and 6 layers), we sweep transformer depth from $L{=}0$ to $L{=}6$ under the same training recipe using \emph{Focused} prompts. As visualized in Fig.~\ref{fig:depth_trend}, a shallow stack already captures most of the benefit: $L{=}3$ attains the best \textsc{MAG}, \textsc{ENV}, and \textsc{KLD} (9.99, 3.41, 10.95), while $L{=}0$ yields the lowest $\Delta\text{IB}$ (0.82) and $L{\in}\{5,6\}$ achieve the best W-dis distance (0.77). Depth increases do not improve all metrics monotonically ----notably, $L{=}4$ underperforms on \textsc{KLD} (12.25)---but we observe a mild polishing of distributional alignment (W-dis) as depth grows from 3 to 6. Training converges within $\sim$175--198 epochs across all depths, suggesting diminishing returns beyond $L{=}3$ once prompts inject salient, frame-focused semantics. In general, $L{=}3$ offers the best balanced performance, with $L{=}5$/$6$ providing small gains in W-dis at the cost of weaker \textsc{KLD}. 
\emph{Injecting focused, frame-level semantics via prompt design reduces the need for deep self-attention: three layers suffice for strong overall quality, while additional depth mostly offers minor distributional polish.}

\section{Conclusion} Using LVLM-derived textual cues as visual context, we disentangle which semantics matter for remixing and find that \textit{camera focus} and \textit{scene background} most reliably improve the quality of perceptual mixes. Our SemMix achieves better results compared to VisAH with fewer parameters and shallow transformers, showing that precise visual semantics, not heavy video encoders, are the key to visually aligned audio remixing.

\bibliographystyle{IEEEbib}
\bibliography{refs}

\end{document}